\documentclass[11pt,a4paper]{article}
\usepackage[T1]{fontenc}
\usepackage[utf8]{inputenc}
\usepackage{authblk}
\usepackage{amsmath,amssymb,epsfig,cite,dsfont}
\usepackage{array,multirow}
\usepackage {color}
\usepackage[toc,page]{appendix}
\numberwithin{equation}{section}

\usepackage[colorlinks=true
,urlcolor=blue
,anchorcolor=blue
,citecolor=blue
,filecolor=blue
,linkcolor=blue
,menucolor=blue
,pagecolor=blue
,linktocpage=true
,pdfproducer=medialab
,pdfa=true
]{hyperref}

\advance\hoffset by -1.1truecm
\advance\voffset by -1.0truecm
\newcommand{\be}{\begin{equation}}
\newcommand{\ee}{\end{equation}}
\newcommand{\bea}{\begin{eqnarray}}
\newcommand{\eea}{\end{eqnarray}}

\usepackage{mathrsfs}
\usepackage{authblk}

\numberwithin{equation}{section}
\usepackage{float}
\restylefloat{figure}
\newcounter{appendice}

\begin{document}

\title{
\vspace{2.2cm}\begin{flushleft} \bf{BRST Symmetry: Boundary Conditions and Edge States in QED
\linethickness{.05cm}\line(1,0){433}
}\end{flushleft}}
\author[1]{Nirmalendu~Acharyya \thanks{Nirmalendu.Acharyya@ulb.ac.be}}
\author[2,3]{A.~P.~Balachandran \thanks{balachandran38@gmail.com}}
\author[4]{ Ver\'{o}nica~Errasti~D\'{i}ez \thanks{vediez@physics.mcgill.ca}}
\author[5]{ P.~N.~Bala~Subramanian \thanks{pnbalasubramanian@gmail.com}}
\author[5]{Sachindeo~Vaidya \thanks{vaidya@cts.iisc.ernet.in}}
\affil[1]{\small Optique Nonlin\'eaire Th\'eorique, Universit\'e libre de Bruxelles (U.L.B.), CP 231, Belgium}
\affil[2]{\small Physics Department, Syracuse University, Syracuse, New York 13244-1130, U.S.A.}
\affil[3]{\small Institute of Mathematical Sciences, C.I.T Campus, Chennai, TN 600113, India}
\affil[4]{\small Physics Department, McGill University, 3600 University St., Montreal, QC H3A 2T8, Canada}
\affil[5]{\small Centre for High Energy Physics,  Indian Institute of Science, Bangalore-560012,
India}
\date{}

\renewcommand\Authands{ and }

\maketitle

\begin{abstract}
In manifolds with spatial boundary, BRST formalism can be used to quantize gauge theories. We show that,
in a $U(1)$ gauge theory, only a
subset of all the boundary conditions allowed by the self-adjointness of the Hamiltonian preserves BRST symmetry.
Hence, the theory can
be quantized using BRST formalism only when that subset of boundary conditions is considered. We also show that for
such boundary conditions,
there exist fermionic states which are localized near the boundary.
\end{abstract}

\section{Introduction}

Topological insulators and their surface modes are subjects of emerging interest
(for example, see~\cite{Fu:2008zzb,Fidkowski,Gurarie,Cheng:2012xh,XLQi}).
Especially, understanding a two-dimensional topological insulator in the light of fractional Hall effect
has been a priority in the subject for the last decade~\cite{levin,Maciejko}. 
In this context, theories of gauge fields interacting with matter, especially in
two and three spatial dimensions,
have gained importance.
We investigate the quantization of $U(1)$ gauge theories  with Dirac fermions from this perspective. 

Quantization of gauge theories using BRST
formalism is conventional.  It is elegant, yet simple. One introduces a ghost field for every constraint of the system.
This breaks the gauge
symmetry, but introduces a new global symmetry (called BRST symmetry) generated by appropriate combinations of
the ghosts and the constraints.
The generators of this new global symmetry
are fermionic and hence nilpotent and the physical Hilbert space is identified by its cohomology. 

We are interested in systems like topological insulators. All such real systems available for experiment are of finite
size and hence have spatial boundaries. The presence of boundaries, in general, can reduce the symmetry of the system.  As a reflection of this, 
all boundary conditions
might not preserve the symmetry (as shown in~\cite{nirmalendu}). 
Therefore, boundary conditions naturally assume significance in the discussion of gauge theories
in manifolds with boundaries  and their
quantization using the BRST formalism. 

The boundary conditions cannot be chosen arbitrarily. Rather we need to only impose those boundary
conditions which define domains of self-adjointness of the Hamiltonian~\cite{bal1,aim,bal2}. However, all such 
domains might not be preserved under
BRST transformations. Therefore, in order to quantize the system by BRST formalism, we must choose only those
boundary conditions which not only define a self-adjoint  Hamiltonian, but are also consistent with the
BRST symmetry. 

The presence of boundaries also naturally leads to the discussion of edge states, which, if extant, play
an important role in the physics of the boundary in systems like topological insulators~\cite{XLQi,cheng2}. 

In this paper, we consider a $U(1)$ gauge theory with Dirac fermions on a $(d+1)$-dimensional  manifold $M\times\mathbb{R}$ with
spatial boundary $\partial M$ { of codimension one}. In section \ref{YMDsystem}, we review the usual discussion of a
$U(1)$ gauge theory.
In section \ref{BRSYsym}, we introduce the
ghosts and invoke BRST symmetry.  We obtain the set of all allowed boundary conditions by demanding the self-adjointness of
the gauge
fixed Hamiltonian.  
We show that, out of the  set of boundary conditions on the gauge fields
consistent with the self-adjointness of the Hamiltonian, only some of
them are invariant under
BRST transformations. This subset of boundary conditions is the same as that
obtained by quantization  of the system
using the canonical formalism~\cite{{bal1},{Asorey:2015sra}}. 

However, we show that there is no such constraint on the  boundary conditions of the Dirac fermions.
Hence, any domain of self-adjointness of the Dirac Hamiltonian is compatible with the BRST symmetry. 

For a system like a topological insulator, we are further required to use physical conditions to choose the suitable
boundary conditions from this set of allowed BRST-preserving boundary conditions. 

Finally, we discuss the possibility of fermionic edge states in the system.
In a simple $(2+1)$-dimensional geometry, we solve for the eigensates of the Hamiltonian in the limit of small 
coupling constant, with boundary conditions  that ensure the self-adjointness of the Hamiltonian
and preserve the BRST symmetry. We show that there
exist fermionic edge states (protected by a mass gap), which interact with soft photons
 and do not break BRST symmetry. These states should be experimentally detectable.

\section{The Maxwell-Dirac System}
\label{YMDsystem}

Consider a gauge theory of Dirac fields  (which we call $U(1)$ Maxwell-Dirac system) on a $(d+1)$-dimensional flat manifold $M\times \mathbb{R}$
with spatial boundary $\partial M$ of codimension one.
We choose the metric $
g^{\mu\nu}=diag\,(1,{-1,\ldots,-1}).$
We use the convention that Greek alphabets $(\mu,\nu,\ldots)$ range from $0$ to $d$ and indices with
Latin alphabets $(i,j,\ldots)$ range from $1$ to $d$.

The $U(1)$  gauge fields $A_\mu$ are Hermitian
\begin{eqnarray}
A_\mu^\dagger = A_\mu
\end{eqnarray}
and the field strength is given by 
\begin{eqnarray}
F_{\mu\nu}=\partial_\mu A_\nu-\partial_\nu A_\mu.
\end{eqnarray}
The covariant derivative is
\begin{eqnarray}
D_\mu=\partial_\mu-ieA_\mu,
\end{eqnarray}
with $e$ the { gauge} coupling constant.

The Maxwell-Dirac  action is given by 
\begin{eqnarray}
&&S=\int_{M\times \mathbb{R}} d^{d+1}x\,\,\mathcal{L}, \\ 
\label{ymlag}
&&\mathcal{L}=-\frac{1}{4}F_{\mu\nu}F^{\mu\nu}+i\bar{\psi}\gamma^\mu D_\mu \psi-m\bar{\psi}\psi,
\end{eqnarray}
where $m$ is the mass of the fermions and $\bar{\psi}=\psi^\dagger\gamma^0$.
The Gamma matrices generate the Clifford algebra:
\begin{eqnarray}
\{\gamma^\mu,\gamma^\nu\}=2g^{\mu\nu}, \quad\quad \gamma^{0\dagger} =\gamma^0, \quad \quad \gamma^{i\dagger}
=-\gamma^i.
\end{eqnarray}

The conjugate momenta to the gauge fields $A_\mu$ and the fermions $\psi,\bar{\psi}$ are given by
\begin{eqnarray}
\label{momenta1}
& \hspace{-0.9cm}\Pi_{\textrm{gauge}}^{i}\equiv \frac{\partial\mathcal{L}}{\partial\dot{A}_i}=F^{i0}, &\quad\quad
\Pi_{\textrm{gauge}}^{0}\equiv \frac{\partial\mathcal{L}}{\partial\dot{A}_0}=0, \\
\label{momenta2}
&\Pi_\psi\equiv \frac{\partial\mathcal{L}}{\partial\dot{\psi}}=i\bar{\psi}\gamma^0=i\psi^\dagger, &\quad\quad
\Pi_{\bar{\psi}}\equiv \frac{\partial\mathcal{L}}{\partial\dot{\bar{\psi}}}=0,
\end{eqnarray}
where the dot denotes derivation with respect to time. 
Notice that the field $A_0$ is not dynamical. In other
words, the Lagrangian (\ref{ymlag}) does not depend on $\dot{A}_0$. As a consequence, the momentum
$\Pi_{\textrm{gauge}}^{0}$ conjugate to $A_0$ vanishes and thus $A_0$ is arbitrary and plays the role
of a Lagrange multiplier. In fact, $\Pi_{\textrm{gauge}}^{0}=0$ is a primary constraint and as such it is part
of the gauge symmetry generator.\footnote{A detailed review of primary constraints and their relation to gauge
transformations can be found in~\cite{Pitts:2013uxa}.}
The Hamiltonian is
\begin{align}
H&=\int_M d^dx\, \left(\Pi^i_{\textrm{gauge}}\dot{A}_i+\Pi_\psi\dot{\psi}-\mathcal{L}\right) \\
&=\int_M d^dx\left[\frac{1}{2}(\Pi^i_{\textrm{gauge}})^2 +\frac{1}{4}F_{ij}F^{ij}
-\Pi_\psi\gamma^0(\gamma^i D_i+im)\psi+G A_0\right],
\end{align}
where $G$, the Gauss law operator, is 
\begin{eqnarray}
\label{wrong}
G=\partial_i \Pi^i_{\textrm{gauge}}-ie\Pi_\psi\psi.
\end{eqnarray}
In order for this operator to generate gauge transformations infinitesimally, the correct expression for the Gauss law is not
(\ref{wrong}), but rather
\begin{eqnarray}
\label{gausslaw}
G(h)\equiv \int_M d^{d}x\,[\Pi^i_{\textrm{gauge}}\partial_i-ie\Pi_\psi\psi] h(x^0,\vec{x})=0,
\end{eqnarray}
with $h(x^0,\vec{x})$ a test function that vanishes at the spatial boundary of our manifold:
\begin{equation}
h(x^0,\vec{x})\Big|_{\partial M}=0.
\end{equation}
The operator $G(h)$ vanishes on quantum state vectors in the physical subspace.

This analysis must followed by a suitable choice of boundary conditions on $A_i$ and $\psi$ invoking the self-adjointness of
the Hamiltonian and subsequent canonical quantization, as in~\cite{bal1}. 

\section{BRST Symmetry}
\label{BRSYsym}
 In this section, we  explore the quantization of the Maxwell-Dirac theory using BRST formalism. 
The  BRST formalism deals with the quantization of gauge fields in a rigorous mathematical framework.
This approach amounts to replacing the gauge symmetry of the theory by a global BRST symmetry, which enlarges the number
of degrees of freedom
in the original theory.  In this enlarged Hilbert space, the usual canonical quantization can be performed.
Then, restricting attention to BRST-invariant states, one recovers the Hilbert
space of physical states of the original theory.

The gauge symmetry of the above Maxwell-Dirac system can be replaced by the BRST global symmetry by introducing
three additional fields: an auxiliary field $\mathcal{B}$, a ghost field
$\mathcal{G}$ and an anti-ghost field $\bar{\mathcal{G}}$. This new action is given by
\begin{eqnarray}
\label{brstac}
&&S_{BRST}=\int_M d^{d+1}x\,\,\mathcal{L}_{BRST}, \\
&&\mathcal{L}_{BRST}=\mathcal{L}+\mathcal{B}(\partial^\mu A_\mu-\frac{\zeta}{2}\mathcal{B})
+(\partial^\mu \bar{\mathcal{G}})(\partial_\mu\mathcal{G}),
\end{eqnarray}
where $\zeta$ is a real parameter and $\mathcal{L}$ is given in (\ref{ymlag}).

In the presence of such new fields, the conjugate momentum $\Pi_{\textrm{gauge}}^{0}$ becomes non-zero:
\begin{equation}
\label{momenta3}
\Pi_{\textrm{gauge}}^{0}=\mathcal{B}.
\end{equation}
On the other hand, the conjugate momenta to the auxiliary, ghost and anti-ghost fields are given by
\begin{eqnarray}
\label{momenta4}
 \Pi_{\mathcal{B}}\equiv \frac{\partial\mathcal{L}_{BRST}}{\partial \dot{\mathcal{B}}}=0, \quad\quad
 \Pi_{\mathcal{G}}\equiv  \frac{\partial\mathcal{L}_{BRST}}{\partial \dot{\mathcal{G}}}=
\dot{\bar{\mathcal{G}}}, \quad\quad
 \Pi_{\bar{\mathcal{G}}}\equiv  \frac{\partial\mathcal{L}_{BRST}}{\partial \dot{\bar{\mathcal{G}}}}=
\dot{{\mathcal{G}}}.
\end{eqnarray}
The Hamiltonian is
\begin{align}
H=&\int_M d^dx\,\left(\Pi^\mu_{\textrm{gauge}}\dot{A}_\mu+\Pi_\psi\dot{\psi}
+\Pi_{\mathcal{G}}\dot{\mathcal{G}}+\dot{\bar{\mathcal{G}}}\Pi_{\bar{\mathcal{G}}}
-\mathcal{L}_{BRST}\right) \\
=&\int_M d^dx\left[\frac{\zeta-1}{2}(\Pi^0_{\textrm{gauge}})^2+\frac{1}{2}(\Pi^0_{\textrm{gauge}}
-\partial^iA_i)^2-\frac{1}{2}(\partial^i
A_i)^2
+\frac{1}{2}(\Pi^i_{\textrm{gauge}}+\partial_i A_0)^2\right. \nonumber \\
&\left.-\frac{1}{2}(\partial_i{A}_0)^2+\frac{1}{4}F_{ij}F^{ij}
-\Pi_\psi\gamma^0(\gamma^i D_i +im-ie\gamma^0 A_0)\psi
+\Pi_{\mathcal{G}}\Pi_{\bar{\mathcal{G}}}-(\partial^i\bar{\mathcal{G}})(\partial_i\mathcal{G})\right].
\end{align}
Defining
\begin{eqnarray}
\mathcal{P}^0\equiv\Pi^0_{\textrm{gauge}}-\partial^i A_i, \quad\quad \mathcal{P}^i\equiv\Pi^i_{\textrm{gauge}}-\partial^i A_0,
\end{eqnarray}
we can rewrite the Hamiltonian as
\begin{align}
H=&\int_M d^dx\left[\frac{\zeta-1}{2}(\Pi^0_{\textrm{gauge}})^2+\frac{1}{2}(\mathcal{P}^0)^2+\frac{1}{2}(\mathcal{P}^i)^2
-\Pi_\psi\gamma^0(\gamma^i D_i+im-ie\gamma^0 A_0)\psi\right. \nonumber \\
&\left.+\Pi_{\mathcal{G}}\Pi_{\bar{\mathcal{G}}}
-\bar{\mathcal{G}}(\partial_i^2\mathcal{G})+\frac{1}{2}A_0(\partial_i^2 A_0)
+A_i(\partial_i\partial_j A_j)-\frac{1}{2}A_i(\partial_j^2 A_i)\right] \nonumber \\
&+\int_{\partial M}d^{d-1}x\left[\bar{\mathcal{G}}\partial_n\mathcal{G}-\frac{1}{2}A_0\partial_n A_0
+\frac{1}{2}A_i\partial_n A_i
-\frac{1}{2}A_n\partial_i A_i-\frac{1}{2}A_i\partial_i A_n\right],
\end{align}
where $n$ denotes the outward pointing unit vector of the boundary $\partial M$.
Removing the boundary terms, the Hamiltonian is 
\begin{align}
H=&\int_M d^dx\left[\frac{\zeta-1}{2}(\Pi^0_{\textrm{gauge}})^2+\frac{1}{2}(\mathcal{P}^0)^2+\frac{1}{2}(\mathcal{P}^i)^2
-\Pi_\psi\gamma^0(\gamma^i D_i+im-ie\gamma^0 A_0)\psi\right. \nonumber \\
&\left.+\Pi_{\mathcal{G}}\Pi_{\bar{\mathcal{G}}}
-\bar{\mathcal{G}}(\partial_i^2\mathcal{G})+\frac{1}{2}A_0(\partial_i^2 A_0)
+\frac{1}{2}A_i(-\partial_j^2 A_i+2\partial_i\partial_j A_j)\right].
\end{align}

The fields  can be expanded in the basis of the eigenfunctions of the following operators:
\begin{eqnarray}
\hspace*{-3cm}&\quad\quad\quad\quad\quad\quad-\partial_j^2A_i+2\partial_i\partial_jA_j=\omega^2A_i,
&\quad\quad \partial_i^2A_0=
\omega_0^2A_0, \nonumber \\
\hspace*{-3cm}&\partial_i^2\mathcal{G}=\omega_g^2\mathcal{G}, &\quad\quad H_D\psi=E_D\psi, \label{ops}
\end{eqnarray}
where $H_D$ is the Dirac Hamiltonian given by
\begin{eqnarray}
H_D=i\gamma^0\gamma^\mu D_\mu-m\gamma^0
\end{eqnarray}
and $\omega^2,\omega_0^2,\omega_g^2,E_D\geq0$, by the requirement of positivity of the Hamiltonian.

As we show in detail in the appendix , this requirement leads to the following most general  boundary
 conditions on the fields:
\begin{eqnarray}
\label{BCperp}
&& (\vec{A}_\perp+i\vec{F}_{n\perp})(x)\Big|_{\partial M}=U_\perp(x)(\vec{A}_\perp-i\vec{F}_{n\perp})(x)\Big|_{\partial M}, \\
\label{BCn}
&& (A_n+i\partial_i A_i)(x)\Big|_{\partial M}=U_n(x)(A_n-i\partial_i A_i)(x)\Big|_{\partial M}, \\
\label{BC0}
&& (A_0+i\partial_n A_0)(x)\Big|_{\partial M}=U_0(x)(A_0-i\partial_n A_0)(x)\Big|_{\partial M}, \\
\label{BCg}
&& (\mathcal{G}+i\partial_n\mathcal{G})(x)\Big|_{\partial M}=U_g(x)(\mathcal{G}-i\partial_n\mathcal{G})(x)\Big|_{\partial M}, \\
\label{BCpsi}
&& \psi_+(x)\Big|_{\partial M}=U_F(x)\gamma^0\psi_-(x)\Big|_{\partial M}.
\end{eqnarray}
Here, $\forall x\in\partial M$, we have defined
\begin{eqnarray}
\label{psipm}
\psi_\pm\equiv\frac{1}{2}(\mathbb{I}\pm\gamma^0\vec{\gamma}\cdot\hat{n})\psi, \quad\quad F^{(A)}_{in}\equiv\partial_iA_n
-\partial_nA_i
\end{eqnarray}
and the operators $U_\perp, U_n, U_0, U_g$  and $U_F$ satisfy 
\begin{eqnarray}
\begin{array}{cc}
U_\perp^\dagger U_\perp =\mathbb{I},\quad \quad  U_n^\dagger U_n=\mathbb{I}, \quad \quad  U_0^\dagger U_0=\mathbb{I}, \\  \\
U_g^\dagger U_g=\mathbb{I}, \quad \quad   U_F^\dagger U_F=\mathbb{I}, \quad \quad  [U_F,\gamma^0\vec{\gamma}\cdot\hat{n}]=0.
\end{array}
\end{eqnarray}

The ghost field can be expanded in a complete orthonormal set of functions $\{H_k(x^0,x^i)\}$ as
\begin{eqnarray}
\label{ghostexpansion}
\mathcal{G}(x^0,x^i)=\sum_k\mathcal{C}_kH_k(x^0,x^i).
\end{eqnarray}
Using the Gauss law (\ref{gausslaw}), the momenta (\ref{momenta1})-(\ref{momenta2}) and
(\ref{momenta3})-(\ref{momenta4}) and the above ghost field expansion, the BRST charge can be written as
\begin{eqnarray}
\label{brstcharge}
\hat{\Omega}\equiv  G(\sum_k\mathcal{C}_kH_k)-i\int_M d^dx\,\,\Pi_{\bar{\mathcal{G}}}\Pi_{\textrm{gauge}}^{0},
\end{eqnarray}
where
\begin{eqnarray}
G(\sum_k\mathcal{C}_kH_k)\equiv \int_M\,\,[\Pi_{\textrm{gauge}}^{i}(\partial_i
-ie\Pi_\psi\psi]\sum_k\mathcal{C}_k H_m(x^0,x^i).
\end{eqnarray}

This BRST charge generates the variation of the fields under which the action (\ref{brstac}) remains invariant.
In this work we are only interested in the BRST variation of the gauge fields $A^{i}$
and fermions $\psi$.
Upon imposing the following canonical commutation relations:
\begin{eqnarray}
&&[\Pi_{\textrm{gauge}}^{i}(x^0,\vec{x}),A^{j}(x^0,\vec{y})]=-i\delta^{ij}\delta^d(\vec{x}-\vec{y}),\\
&&\{\Pi_\psi(x^0,\vec{x}),\psi(x^0,\vec{y})\}=\delta^d(\vec{x}-\vec{y}),
\end{eqnarray}
the BRST variations of our interest are
\begin{eqnarray}
\label{deltaA0}
&&\delta A^{0}=i\epsilon[\hat{\Omega},A^{0}]=
\epsilon\partial^{0}\mathcal{G}, \\
\label{deltaA}
&&\delta A^{i}=i\epsilon[\hat{\Omega},A^{i}]=
\epsilon\partial^{i}\mathcal{G}, \\
\label{deltapsi}
&&\delta\psi=i\epsilon\{\hat{\Omega},\psi\}=-\epsilon\mathcal{G}\psi,\\
\label{deltag}
&&\delta\mathcal{G}= i\epsilon\{\hat{\Omega}, \mathcal{G}\} = 0,
\end{eqnarray}
where $\epsilon$ is a Grassmannian number.

\section{The Boundary Conditions}
\label{BCs}

The boundary conditions (\ref{BCperp})-(\ref{BCpsi}) which preserve the self-adjointness of the Hamiltonian are not
consistent with BRST symmetry. In the following we show that only a smaller subset of these boundary conditions preserve
BRST symmetry.

As we mentioned, the BRST charge $\hat{\Omega}$ in (\ref{brstcharge}) generates a global BRST
symmetry in the action (\ref{brstac}).
However, in order for $\hat{\Omega}$ to generate the BRST symmetry infinitesimally, all $H_k(x^0,x^i)$ in 
(\ref{ghostexpansion}) must vanish on $\partial M$:
\begin{eqnarray}\label{bcghost11}
\mathcal{G}\Big|_{\partial M}=\sum_k\mathcal{C}_kH_k(x^0,x^i)\Big|_{\partial M}=0.
\end{eqnarray}
This requirement implies that
\begin{eqnarray}
\label{bcghost}
\vec{\nabla}_\perp\mathcal{G}(x)\Big|_{\partial M}=0, \quad\quad \partial_0{\mathcal{G}}(x)\Big|_{\partial M}=0.
\end{eqnarray}
Thus, the BRST transformation (\ref{deltag}) enforces $U_g=-\mathbb{I}$ in (\ref{BCg}). 

\subsection{Allowed boundary conditions on $A_\mu$}

From (\ref{deltaA0}) and (\ref{bcghost}) it follows that
\begin{eqnarray}
\label{deltaFnperp}
\delta A_0(x)\Big|_{\partial M}=0.
\end{eqnarray}
Using the above in (\ref{BC0}), we get 
\begin{eqnarray}
[1+U_0(x)]\delta (\partial_nA_0)(x)\Big|_{\partial M}= \epsilon[1+U_0(x)] \partial_0 \partial_n
\mathcal{G}(x)\Big|_{\partial M}=0, 
\end{eqnarray} 
As $\partial_n \mathcal{G}(x)\Big|_{\partial M}\neq 0$ in general, the above implies that BRST symmetry enforces
$U_0=-\mathbb{I}$ and hence,  the BRST-preserving boundary condition on $A_0$ is
\begin{eqnarray}
\label{gaugebc}
A_0(x)\Big|_{\partial M}=0.
\end{eqnarray}
For any other boundary condition on $A_0$, the BRST symmetry will be broken. 

Form (\ref{deltaA}) and (\ref{bcghost}) it is easy to check that 
\begin{eqnarray}
\label{deltaFnperp1}
 \delta \vec{A}_\perp(x)\Big|_{\partial M}=0, \quad\quad
\delta\vec{F}_{n\perp}(x)\Big|_{\partial M}
=0.
\end{eqnarray}
Consequently, the BRST variation of  (\ref{BCperp}) becomes trivial and the boundary conditions (\ref{BCperp})
are allowed by BRST symmetry for all $U_\perp$. 
In a similar fashion, the boundary conditions (\ref{BCn}) are also not constrained by the
BRST symmetry  and any  $U_n(x)$ is allowed.

These are the same set of boundary conditions (\ref{gaugebc}) and (\ref{BCperp})
that one obtains if the theory is quantized using Dirac constraints in canonical
formalism~\cite{{bal1},{Asorey:2015sra}}. 

In a system like a topological insulator  where boundaries play a vital role, we can further use physical conditions
to constrain this allowed set of BRST-preserving boundary conditions. The surface of a topological
insulator, unlike the bulk (which is an insulator), behaves like a conductor.
Therefore, the tangential component of the electric field must vanish on the boundary of the topological
insulator. Then,
recalling that $A_0$ vanishes on the boundary, we need to choose
\begin{eqnarray}
\vec{A}_\perp(x)\Big|_{\partial M}=0. 
\end{eqnarray}
This is one  of the allowed boundary conditions from the set (\ref{BCperp})
(for this case, $U_\perp = - \mathbb{I}$) and this ensures that  the tangencial component of the electric field
$\vec{E}_\perp = \partial_0 \vec{A}_\perp -\vec{\nabla}_\perp A_0$
vanishes on the boundary. Also, this is one of the boundary conditions obtained in \cite{Asorey:2015sra} using canonical formalism. 
\subsection{Fermionic Boundary {Conditions}}

From (\ref{deltapsi}) and  (\ref{bcghost11}) it follows that
\begin{eqnarray}
\delta\psi(x)\Big|_{\partial M}=0.
\end{eqnarray}
Using the above in (\ref{psipm}), it is easy check that
\begin{equation}
\delta \psi_\pm (x)\Big|_{\partial M} =0. 
\end{equation}
A BRST variation of the boundary condition (\ref{BCpsi}) is thus trivial and hence, the boundary
condition (\ref{BCpsi}) is compatible with the BRST symmetry for any choice of $U_F$. 
Thus, the BRST symmetry constrains the boundary conditions on the gauge fields $A_\mu$, but it
does not constrain the fermionic boundary conditions.

\section{A $(2+1)$-Dimensional Example}
\label{3d}

In the following, we consider the $(2+1)$-dimensional case, which is particularly relevant in the context of
topological insulators.
Consider the $(2+1)$-dimensional manifold
\begin{eqnarray}
\tilde{M}\equiv\{x^0,x^1,x^2: \quad x^1\leq 0\}
\end{eqnarray}
with spatial boundary
\begin{eqnarray}
\partial \tilde{M}=\{x^0,x^1,x^2: \quad x^1= 0\}.
\end{eqnarray}
We choose the following representation of the Gamma matrices:
\begin{eqnarray}
\gamma^0=\sigma^2, \quad\quad \gamma^1=i\sigma^1, \quad\quad \gamma^2=i\sigma^3,
\end{eqnarray}
with $\sigma^i$'s the Pauli matrices. It follows then that $\psi_\pm$ are given by
\begin{eqnarray}
\psi_+=\left(
\begin{array}{ccc}
\psi_1 \\
0
\end{array}
\right), \quad\quad \psi_-=\left(
\begin{array}{ccc}
0 \\
\psi_2
\end{array}
\right).
\end{eqnarray}
It is easy to check that the matrix $U_F$ must then take the form
\begin{eqnarray}
U_F=\left(
\begin{array}{ccc}
e^{i\theta} & 0 \\
0 & e^{i\tilde{\theta}}
\end{array}
\right), \quad\quad \theta,\tilde{\theta}\in\mathbb{R}.
\end{eqnarray}
The boundary conditions in (\ref{BCpsi}) in this case are simply
\begin{eqnarray}
\label{2dbcfermion}
\psi_1\Big|_{x_1=0}=-ie^{i\tilde{\theta}}\psi_2\Big|_{x_1=0}
\end{eqnarray}
and the gauge fields satisfy the following boundary conditions:
\begin{eqnarray}
A_0\Big|_{x_1=0}=0, \quad\quad A_2\Big|_{x_1=0}=0.
\end{eqnarray}
The vanishing of $A_0$ on the boundary $x_1=0$  is required by BRST symmetry. However, the condition $ A_2\Big|_{x_1=0}=0$
is one of the many boundary conditions (\ref{BCperp}) that preserves BRST symmetry. We choose this particular
boundary condition because it leads to the vanishing of
the tangential component of the electric field  on the boundary, as it should in  a topological insulator.

It is easy to check that
\begin{eqnarray}
A_0^{(k)}=0, \quad\quad A_1^{(k)}=a_k k_2\cos(k_1x_1)\cos(k_2x_2), \quad\quad A_2^{(k)}=a_k k_1\sin(k_1x_1)\sin(k_2x_2),
\end{eqnarray}
 with $a_k\in\mathbb{C}$, satisfy the above boundary conditions and are solutions of the eigenvalue equations
\begin{eqnarray}
-\partial_j^2A_i^{(k)}+2\partial_i\partial_jA_j^{(k)}=\omega_k^2 A_i^{(k)},
\quad\quad \partial_i^2 A_0^{(k)}=\omega_0^2A_0^{(k)},
\end{eqnarray}
with
\begin{eqnarray}
\omega_k^2=k_1^2+k_2^2, \quad\quad \omega_0=0.
\end{eqnarray}
Thus, the gauge field can be expressed as
\begin{eqnarray}
\label{gafield}
A_0=0, \quad A_1=\sum_{k_1,k_2}a_k k_2\cos(k_1x_1)\cos(k_2x_2), \quad A_2=\sum_{k_1,k_2}a_k k_1\sin(k_1x_1)\sin(k_2x_2).
\end{eqnarray}
Demanding reality of the gauge fields yields
\begin{eqnarray}
a_k^\ast = a_k {\quad\implies\quad a_k\in\mathbb{R}}.  
\end{eqnarray}

The ghost field can be expanded in the eigenfunctions of the scalar Laplacian
\begin{eqnarray}
H_{\tilde{k}}=e^{i\tilde{k}_2x_2}\sin(\tilde{k}_1x_1)
\end{eqnarray}
with eigenvalues
\begin{eqnarray}
\omega_g^2=\tilde{k}_1^2+\tilde{k}_2^2.
\end{eqnarray}
Hence, the ghost field can be expressed as
\begin{eqnarray}
\mathcal{G}=\sum_{\tilde{k}_1,\tilde{k}_2}\mathcal{C}_{\tilde{k}}H_{\tilde{k}}, \quad\quad \mathcal{C}_{\tilde{k}}\in\mathbb{C}.
\end{eqnarray}

\subsection{Eigenstates of the Dirac Operator }
\label{edgest}

In this section, we solve for the fermionic edge states in $\tilde{M}$ when the coupling constant $g$ is small.
 We want to consider the interaction of the fermions with photons of very small energies.
For such soft-photons, we can terminate the sums in (\ref{gafield}) at small values of $k_1,k_2$,
which in turn imply a small $\omega_k$.

For simplicity, we will assume that $\tilde{\theta}=\pi/2$ in (\ref{2dbcfermion}).
With this choice, the fermionic boundary condition (\ref{2dbcfermion}) reduces to
\begin{eqnarray}
\label{nfbc}
\psi_1\Big|_{x_1=0}=\psi_2\Big|_{x_1=0}.
\end{eqnarray}
However, it is not difficult to
generalize the analysis to arbitrary
$\tilde{\theta}$. 

For small gauge coupling constant $e$, we  expand the field
$\psi$ in {$e$} as
\begin{eqnarray}
&& \psi=\chi+{e}\xi+\ldots 
\end{eqnarray}
The eigenvalue equation for  the Dirac fermions
\begin{eqnarray}\label{eigeneqn_1}
H_D\psi\equiv [i\gamma^0\gamma^i(\partial_i-i{e}A_i)+{e}A_0+m\gamma^0]\psi=E\psi,
\quad\quad E\in\mathbb{R},
\end{eqnarray}
at order 1, leads to 
\begin{eqnarray}
i\gamma^0(\gamma^i\partial_i-im)\chi=E\chi,
\end{eqnarray}
subject to the boundary condition (\ref{nfbc}). It is easy to see that the above has solution
\begin{eqnarray}
\chi=\left(
\begin{array}{ccc}
1 \\
1
\end{array}
\right)e^{mx_1+iEx_2}.
\end{eqnarray}
At order {$e$}, the eigenvalue equation (\ref{eigeneqn_1}) gives
\begin{eqnarray}
\label{eigeng}
i\gamma^0(\gamma^i\partial_i-im)\xi+\gamma^0\gamma^iA_i\chi=E\xi.
\end{eqnarray}
To solve this, we start by rewriting $A_i$ as
\begin{eqnarray}
&& A_1=\sum_{k_1,k_2}\frac{a_k}{4}k_2(e^{ik_1x_1}+e^{-ik_1x_1})(e^{ik_2x_2}+e^{-ik_2x_2}), \\
&& A_2=-\sum_{k_1,k_2}\frac{a_k}{4}k_1(e^{ik_1x_1}-e^{-ik_1x_1})(e^{ik_2x_2}-e^{-ik_2x_2}).
\end{eqnarray}
Inserting the ansatz
\begin{align}
\xi=\sum_{k_1,k_2}&\left(\xi_k^{(1)}e^{(ik_1+m)x_1+i(k_2+E)x_2}
+\xi_k^{(2)}e^{(-ik_1+m)x_1+i(k_2+E)x_2}\right. \nonumber \\
&\left.+\xi_k^{(3)}e^{(ik_1+m)x_1+i(-k_2+E)x_2}
+\xi_k^{(4)}e^{(-ik_1+m)x_1+i(-k_2+E)x_2}\right)
\end{align}
in (\ref{eigeng}), we obtain
\begin{eqnarray}
\begin{array}{llll}
&& \xi_k^{(1)}=-\frac{a_k}{4}(2Ek_2-2imk_1+\omega_k^2)^{-1}\left(
\begin{array}{ccc}
2Ek_1+2imk_2-\omega_k^2 \\ 
2Ek_1+2imk_2+\omega_k^2
\end{array}
\right), \\ \\
&& \xi_k^{(2)}=\frac{a_k}{4}(2Ek_2+2imk_1+\omega_k^2)^{-1}\left(
\begin{array}{ccc}
2Ek_1-2imk_2+\omega_k^2 \\ 
2Ek_1-2imk_2-\omega_k^2
\end{array}
\right), \\ \\
&& \xi_k^{(3)}=-\frac{a_k}{4}(2Ek_2+2imk_1-\omega_k^2)^{-1}\left(
\begin{array}{ccc}
2Ek_1-2imk_2-\omega_k^2 \\
2Ek_1-2imk_2+\omega_k^2
\end{array}
\right), \\ \\
&& \xi_k^{(4)}=\frac{a_k}{4}(2Ek_2-2imk_1-\omega_k^2)^{-1}\left(
\begin{array}{ccc}
2Ek_1+2imk_2+\omega_k^2 \\
2Ek_1+2imk_2-\omega_k^2
\end{array}
\right).
\end{array}
\end{eqnarray}
When $\omega_k$ is very small, we can set $\omega_k^2\approx 0$ and hence the above reduces to
\begin{eqnarray}
\begin{array}{llll}
& \xi_k^{(1)}=-\frac{a_k}{4}\frac{Ek_1+imk_2}{Ek_2-imk_1}\left(
\begin{array}{ccc}
1 \\ 
1
\end{array}
\right), \quad\quad
& \xi_k^{(2)}=\frac{a_k}{4}\frac{Ek_1-imk_2}{Ek_2+imk_1}\left(
\begin{array}{ccc}
1\\ 
1
\end{array}
\right), \\ \\
& \xi_k^{(3)}=-\frac{a_k}{4}\frac{Ek_1-imk_2}{Ek_2+imk_1}\left(
\begin{array}{ccc}
1 \\
1
\end{array}
\right), \quad\quad
& \xi_k^{(4)}=\frac{a_k}{4}\frac{Ek_1+imk_2}{Ek_2-imk_1}\left(
\begin{array}{ccc}
1 \\
1
\end{array}
\right).
\end{array}
\end{eqnarray}
Therefore, in the presence of soft photons,
\begin{eqnarray}
\psi= \Big[(e^{mx_1+iEx_2} + e\sum_{k_1,k_2} \left(a_k^{(1)}e^{(ik_1+m)x_1+i(k_2+E)x_2}
+a_k^{(2)}e^{(-ik_1+m)x_1+i(k_2+E)x_2}\right. \nonumber \\
\left.+a_k^{(3)}e^{(ik_1+m)x_1+i(-k_2+E)x_2}
+a_k^{(4)}e^{(-ik_1+m)x_1+i(-k_2+E)x_2}\right)\Big]\left(
\begin{array}{ccc}
1 \\
1
\end{array}
\right)+O(e^2)
\end{eqnarray}
with \begin{eqnarray}
\begin{array}{llll}
 a_k^{(1)}=-\frac{a_k}{4}\frac{Ek_1+imk_2}{Ek_2-imk_1}, \quad\quad 
& a_k^{(2)}=\frac{a_k}{4}\frac{Ek_1-imk_2}{Ek_2+imk_1}, \\ \\
 a _k^{(3)}=-\frac{a_k}{4}\frac{Ek_1-imk_2}{Ek_2+imk_1},\quad\quad 
& a_k^{(4)}=\frac{a_k}{4}\frac{Ek_1+imk_2}{Ek_2-imk_1},
\end{array}
\end{eqnarray}
are eigenmodes of  (\ref{eigeneqn_1})  and satisfy the boundary condition (\ref{nfbc}). 

For a sufficiently large mass $m$, these eigenmodes 
are exponentially damped in the bulk and are localized near the edge $x_1=0$. 
In real systems, like topological insulators, these modes are experimentally detectable.

\section{Discussions}

The BRST formalism provides a natural framework to quantize gauge theories in the presence of spatial boundaries,
which are particularly
important in real systems, like topological insulators. We have shown that in a $U(1)$ gauge theory, out of the set of
all local
boundary conditions on the gauge fields allowed by the self-adjointness of the Hamiltonian, only some  preserve BRST symmetry.
These BRST-preserving boundary conditions are, in general, consistent with observations in a topological insulator. 

The presence of fermionic edge states in the theory is also very interesting from the perspective of a system like a
topological insulator.
These edge states are expected to assume an important role in the physics at the boundary:  it is possible to experimentally
verify
the presence of these fermions localized at the boundary.

To demonstrate the presence of  edge states,
in the previous
section we have
considered a very simple (2+1)-dimensional system with flat boundaries. However, those results can be easily extended
to any spacetime
dimension and to any curved boundary of codimension one.  Also, we considered the fermions to be massive  so that
the edge states are protected by
the corresponding mass gap.  However, one might also consider a
gapless system with time-reversal symmetry. There also,
we expect to find  edge-localized fermions in a similar fashion, though the details in that case will be a bit different.  

\subsection*{ Acknowledgements} 

 V.E.D. is grateful to NSERC (Grant No. 210381) and Keshav
Dasgupta for financial support. V.E.D. thanks the Centre for High Energy Physics, Indian
Institute Science, Bangalore and especially Sachindeo Vaidya for hospitality during
the course of this work. 
\appendix
\section*{Appendices}

\section{Boundary Conditions of the Gauge Fields}

As mentioned in section \ref{BRSYsym}, the fields $A_i$ can be expanded in the basis of the eigenfunctions of the operator
$\hat{\mathcal{O}}\equiv(-\partial_j^2+2\partial_i\partial_j)$. This operator is studied in \cite{Asorey:2015sra}.  To find the domain of self-adjointness of this operator we impose that
\begin{eqnarray}\label{app_eqn1}
\int_M d^dx\,\, \left[B_i^\dagger(-\partial_j^2A_i+2\partial_i\partial_jA_j)-
(-\partial_j^2B_i^\dagger+2\partial_i\partial_jB_j^\dagger)A_i
\right], \quad\quad \forall \,\, A_i\in\mathcal{D}_{\hat{\mathcal{O}}}, \,\, B_i\in\mathcal{D}_{\hat{\mathcal{O}}^\dagger}
\end{eqnarray}
vanishes if and only if the same boundary conditions are imposed on both $A_i$ and $B_i$.   Now (\ref{app_eqn1}) leads to the boundary term
\begin{eqnarray}
\int_{\partial M}d^{d-1}x\,\,\left[B_i^\dagger(-\partial_nA_i+\partial_iA_n)+B_n^\dagger(\partial_iA_i)-
(-\partial_nB_i^\dagger+\partial_i
B_n^\dagger)A_i-(\partial_iB_i^\dagger)A_n\right],
\end{eqnarray}
which must vanish with the same conditions on $A_i$ and $B_i$. The most general local boundary conditions for
which the above rule is satisfied are
\begin{eqnarray}
&&(\vec{A}_\perp+i\vec{F}_{n\perp})(x)\Big|_{\partial M}=U_\perp(x)(\vec{A}_\perp-i\vec{F}_{n\perp})(x)\Big|_{\partial M}, \\
&&(A_n+i\partial_iA_i)(x)\Big|_{\partial M}=U_n(x)(A_n-i\partial_iA_i)(x)\Big|_{\partial M}, \quad\quad x\in \partial M,
\end{eqnarray}
with
\begin{eqnarray}
\vec{F}_{n\perp}=\partial_n\vec{A}_\perp-\vec{\nabla}_\perp A_n, \quad\quad U_\perp^\dagger U_\perp=\mathbb{I}=U_n^\dagger U_n.
\end{eqnarray}

Similarly, $A_0$ can be expanded in the eigenfunctions of $\hat{\mathcal{O}}_0\equiv \partial_j^2$. The domain of self-adjointness of
$\hat{\mathcal{O}}_0$ is obtained by demanding that
\begin{eqnarray}
\int_M d^dx\,\, \left[B_0^\dagger(\partial_j^2A_0)-(\partial_j^2B_0^\dagger)A_0\right], \quad\quad \forall\,\,
A_0\in\mathcal{D}_{\hat{\mathcal{O}}_0},
\,\, B_0\in\mathcal{D}_{\hat{\mathcal{O}}_0^\dagger}
\end{eqnarray}
vanishes with the same boundary conditions on $A_0$ and $B_0$. The above leads to the boundary term
\begin{eqnarray}
\int_{\partial M}d^{d-1}x\,\,\left[B_0^\dagger(\partial_n A_0)-(\partial_n B_0^\dagger)A_0\right],
\end{eqnarray}
which must vanish with the same conditions on $A_0$ and $B_0$. It is easy to check that the most
general local boundary condition which
satisfies the above {requirement} is
\begin{eqnarray}
(A_0+i\partial_n A_0)(x)\Big|_{\partial M}=U_0(x)(A_0-i\partial_nA_0)(x)\Big|_{\partial M}, \quad\quad x\in\partial M,
\end{eqnarray}
with $U_0^\dagger U_0=\mathbb{I}$.

\section{Fermionic Boundary Conditions}

The conventional way to quantize the fermionic field is to expand it in the basis of eigenfunctions of the Dirac
Hamiltonian $H_D$ given by 
\begin{eqnarray}
H_D&=&i\gamma^0\gamma^\mu D_\mu+m\gamma^0 \\
&=&i\gamma^0\gamma^i(\partial_i-ieA_i)+eA_0+\gamma^0m.
\end{eqnarray}
The domain of self-adjointness of $H_D$ can be obtained by demanding that
\begin{eqnarray}
\label{intdom}
\int_M d^dx\,\,\chi^\dagger H_D\psi-\int_Md^dx\,\,(H_D\chi)^\dagger\psi=0, \quad\quad \forall \psi\in\mathcal{D}_{H_D}, \quad \chi\in\mathcal{D}_{H_D^\dagger}
\end{eqnarray}
if and only if $\psi$ and $\chi$ fulfill the same boundary conditions.

We assume that the photon fields are real:
\begin{eqnarray}
A_\mu^\dagger=A_\mu.
\end{eqnarray}
Then, (\ref{intdom}) reduces to
\begin{eqnarray}
i \int_M d^dx\,\,\left[\chi^\dagger\gamma^0\gamma^i\partial_i\psi+ (\partial_i \chi)^\dagger\gamma^0\gamma^i \psi\right]=0,
\end{eqnarray}
which leads to 
\begin{eqnarray}
\int_{\partial M}d^{d-1}x\,\,\chi^\dagger\gamma^0\vec{\gamma}\cdot \hat{n}\psi=0.
\end{eqnarray}
We define the operators
\begin{eqnarray}
P_\pm\equiv\frac{1}{2}(\mathbb{I}\pm\gamma^0\vec{\gamma}\cdot\hat{n}).
\end{eqnarray}
These are projectors, since they satisfy $(P_\pm)^2=P_\pm$. In terms of these projectors, the above integral can be written as
\begin{eqnarray}
\int_{\partial M}d^{d-1}x\,\,\chi^\dagger(P_+-P_-)\psi=\int_{\partial M}d^{d-1}x\,\,\chi^\dagger(P_+^2-P_-^2)\psi=0, 
\end{eqnarray}
Calling $\psi_\pm\equiv P_\pm\psi$, we can further rewrite the above as
\begin{eqnarray}
\int_{\partial M}d^{d-1}x\,\,\left(\chi_+^\dagger\psi_+-\chi_-^\dagger\psi_-\right)=0.
\end{eqnarray}
This requirement leads to the following domain of self-adjointness of $H_D$:
\begin{eqnarray}
\label{selfadjdomain}
\mathcal{D}_{H_D}=\left\{\psi: \quad \psi_+\Big|_{\partial M}=
U_F\gamma^0\psi_-\Big|_{\partial M}\right\},
\end{eqnarray}
where  the matrix $U_F$ satisfies
\begin{eqnarray}
U_F^\dagger U_F=\mathbb{I}.
\end{eqnarray}
Also, as $P_+\gamma^0=\gamma^0 P_-$ and $P_\pm^2=P_\pm$, $U_F$ must satisfy
\begin{eqnarray}
[U_F,\gamma^0\vec{\gamma}\cdot\hat{n}]=0.
\end{eqnarray}

\end{document}